\documentclass[conference]{IEEEtran}
\usepackage{cite}
\usepackage{amsmath,amssymb,amsfonts}
\usepackage{graphicx}
\usepackage{textcomp}
\usepackage{booktabs}
\usepackage{xcolor}
\usepackage[thinc]{esdiff}
\usepackage{algorithm}
\usepackage{algpseudocode}
\usepackage[normalem]{ulem}

\def\BibTeX{{\rm B\kern-.05em{\sc i\kern-.025em b}\kern-.08em
T\kern-.1667em\lower.7ex\hbox{E}\kern-.125emX}}
\begin{document}

\title{Automated Synthesis of Hardware-implementable Analog Circuits for Constrained Optimization}
\author{
Sachin Khoja, Kamlesh Sawant, Palak Jain, Sairaj Dhople, Jason Poon}

\maketitle

\begin{abstract}
This paper presents an automated software toolchain for synthesizing hardware-implementable analog circuits that solve constrained optimization problems. The proposed toolchain supports nonlinear objective functions with linear and quadratic constraints. It maps optimization variables to capacitor voltages, implementing dynamics that enforce Karush–Kuhn–Tucker conditions using operational amplifiers, resistors, capacitors, diodes, and analog multipliers. From high-level problem descriptions in AMPL or MPS, the toolchain generates a SPICE netlist for the analog circuit, simulates it, and verifies that the solutions converge. The projected settling time of the analog circuit depends on circuit parameters, gain-bandwidth product, and slew-rate limits of operational amplifiers, and leverages the inherent parallelism of analog circuits. The proposed toolchain successfully generates circuits with up to $10{,}000$ variables and demonstrates large scalability improvements, achieving up to a $1{,}000\times$ increase in solvable problem size over prior analog hardware demonstrations. Simulation studies further show that the automatically synthesized circuits converge to optimal solutions, achieving more than a $200\times$ speedup compared to IPOPT, a state-of-the-art digital interior-point solver.

\end{abstract}


\section{Introduction}

Constrained optimization is central to modern computational science, with applications in electronic design automation (EDA), machine learning, real-time logistics planning, operations and control of critical infrastructure, and financial modeling~\cite{Boyd_2004, Taylor-2016, Martins_2022, Xia_2024,Kalinin_2025}. The solution of these problems is largely automated through iterative discrete-time algorithms running on conventional von Neumann architectures. However, as the scale and complexity of problems grow, with Moore's Law slowing down, and the end of Dennard scaling, digital solvers face significant challenges. The computational overhead and power consumption associated with solving large-scale optimization problems are becoming prohibitive, creating a bottleneck for many high-performance applications, such as real-time model predictive control (MPC)~\cite{Zometa_2012,Karamanakos_2020,Adegbege_2024,MPC_2024,Simone_2025}.

Prior art has demonstrated that electronic analog computing is a viable alternative, leveraging the energy-minimizing properties of analog circuits to find optimal solutions. This approach potentially offers faster and more efficient computation than conventional digital solvers, making it particularly attractive for real-time optimization applications, such as MPC~\cite{Kouro_2015, Karamanakos_2020}. The concept of using analog circuits to solve optimization problems dates back to the seminal work on Hopfield networks, which demonstrated how a system's dynamics could be engineered to converge to a minimum of an energy function~\cite{Tank_1986}. This was later extended to nonlinear programming, establishing a direct mapping between mathematical optimization constructs and analog circuit elements~\cite{Kennedy_1987, Kennedy_1988}. Recently, several optimization methods and control algorithms, such as gradient-based dynamics and MPC, have been realized with analog circuits~\cite{Vichik_2015,Wu_2024,Adegbege_2024, MPC_2024,Sawant_2025,Simone_2025}. 

Despite these conceptual advances and proof-of-concept demonstrations, the practical application of analog optimization solvers has been limited by the reliance on the manual design of the networks of analog circuits. This bespoke, labor-intensive process requires domain expertise and is not scalable, restricting the complexity of problems that can be addressed. Recent work has explored solving optimization problems using analog circuits, but most of these studies focus on behavioral modeling of the analog circuits for simulation rather than producing designs intended for physical implementation~\cite{Agarwal_2023_A, Agarwal_2023_B ,Boyd_2024}. To date, to the best of the authors' knowledge, most physical hardware implementations used to solve constrained optimization problems are manually designed and have been limited to small-scale problems, typically around ten variables~\cite{Vichik_2014, Vichik_2015,Levenson_2016,Bruno_2021,Adegbege_2024,Poon_2022,Sawant_2024}. Prior work has also demonstrated rapid prototyping of constrained linear–quadratic optimal control using Field-Programmable Analog Arrays (FPAAs), showing that reconfigurable analog hardware can implement MPC algorithms with microsecond-level solution times. However, these demonstrations target a few tens of variables and require manual mapping of the control dynamics to FPAA blocks, so scalability remains limited~\cite{Skibik_2018,Adegbege_2024}. Broadly, scalability barriers have prevented the widespread adoption of analog computing for optimization, as the manual effort required for circuit design far outweighs the potential performance benefits for all but the smallest of problems.

This paper introduces a software toolchain that automates the synthesis of analog circuits for constrained optimization problems featuring nonlinear objective functions with linear and quadratic constraints. Our approach overcomes the limitations of previous work by providing a software toolchain for generating and verifying analog circuits that solve constrained optimization problems from high-level problem descriptions. Given a problem specification in a standard format such as AMPL or MPS, the proposed toolchain automatically generates a SPICE netlist. This netlist can then be simulated to validate the convergence of the large-scale circuit design to the optimal solution of the optimization problem.

This paper demonstrates the synthesis and verification of analog optimization circuits capable of solving \emph{dense problems} with up to $500$ variables and \emph{sparse problems} with up to $10{,}000$ variables. In this context, ``dense'' refers to optimization problems in which most variables interact with many others, whereas ``sparse'' problems are those in which each variable interacts with only a limited subset of others. These results represent a $1{,}000\times$ increase in solvable problem complexity over prior hardware demonstrations. Simulation studies further show that the automatically synthesized circuits converge to optimal solutions, achieving more than a $200\times$ speedup relative to digital solvers such as Interior Point Optimizer (IPOPT)~\cite{Wachter2006}.

The main contributions of this work are as follows:
\begin{itemize}
\item Introduction of an automated toolchain for synthesizing hardware-implementable analog circuits that solve constrained optimization problems directly from high-level problem descriptions.
\item Demonstration and verification of scalable circuit designs for large-scale problems (up to $10{,}000$ variables), overcoming the scalability limits of prior physical analog circuit implementations.
\end{itemize}

The remainder of this paper is organized as follows. Section~\ref{sec:Methodology} describes the methodology utilized for synthesizing analog circuits for solving constrained optimization problems. Section~\ref{sec:Toolchain Overview} provides an overview of the proposed toolchain. Section~\ref{sec:numerical_results} summarizes the numerical results, and Section~\ref{sec:conclusions} concludes the paper.

\section{Methodology} \label{sec:Methodology}
In this section, we introduce the general constrained optimization problem, outline the methodology for synthesizing analog circuits to solve it, and present an illustrative circuit example. 

\subsection{Mapping Optimization Problems to Analog Circuits}
To map optimization problems to analog circuits, we rely on the foundational principle that variables involved in mathematical optimization formulations can be directly associated with the physical quantities of an analog circuit (i.e., voltages and currents). We consider the following general constrained optimization problem in $N$ variables, $M$ inequality constraints, and $P$ equality constraints:
\begin{subequations}
\label{eq:OP}
\begin{align}
&\underset{x \in \mathbb{R}^N}{\mathrm{minimize}} \quad f(x), \quad \\
&\mathrm{subject~to} \quad g_i(x) \geq 0, \; \, \forall \, i \in \{1, \dots, M\}, \label{eq:inequality_constraint} \\
& \qquad  \qquad \quad     h_j(x) = 0, \, \forall \, j \in \{1, \dots, P\},
\end{align}
\end{subequations}
where $x=\begin{bmatrix}x_1 & \cdots & x_N\end{bmatrix}^\top  \in \mathbb{R}^N$ is the vector of optimization variables and $f:\mathbb{R}^N\to\mathbb{R}$ is the objective function. The constraint function, $g:\mathbb{R}^N\to\mathbb{R}^M$ collects inequality constraints and $h:\mathbb{R}^N\to\mathbb{R}^P$ collects equality constraints. The objective and constraint functions are assumed to be continuously differentiable. Problems with box constraints $x_\mathrm{L} \leq x \leq x_\mathrm{U}$ can be reformulated in the inequality form by expressing them as $x-x_\mathrm{L} \geq 0$ and $x_\mathrm{U}-x \geq 0$.

The Lagrangian $\mathcal{L}(x, \lambda, \mu): \mathbb{R}^N~\times~\mathbb{R}^M~\times~\mathbb{R}^P \to\mathbb{R}$ for this problem is
\begin{align*}
\label{eq:Lagrangian}
\mathcal{L}(x, \lambda, \mu)=f(x) + \lambda^\top g(x)+\mu^\top h(x),
\end{align*}
where $\lambda=\begin{bmatrix}\lambda_1 & \cdots & \lambda_M \end{bmatrix}^\top\in \mathbb{R}^M$ and $\mu=\begin{bmatrix}\mu_1 & \cdots & \mu_P\end{bmatrix}^\top \in \mathbb{R}^P$ are the Lagrange multipliers (dual variables). The gradient of the Lagrangian with respect to the optimization variable is
\begin{equation*}
\nabla_x \mathcal{L}(x,\lambda,\mu)= \nabla f(x) + \sum_{i=1}^{M} \lambda_i \nabla g_i(x) + \sum_{j=1}^{P} \mu_j \nabla h_j(x).
\end{equation*}
Here,
$\nabla f(x) = \tfrac{\partial}{\partial x}f(x) \in \mathbb{R}^N$,
$\nabla g_i(x) = \tfrac{\partial}{\partial x}g_i(x)  \in \mathbb{R}^N$, and
$\nabla h_j(x) = \tfrac{\partial}{\partial x}h_j(x)  \in \mathbb{R}^N$
denote the gradients of the objective function, the $i$-th inequality constraints, and the $j$-th equality constraints, respectively,
all with respect to $x$.

The analog circuit implements the following dynamics, which drive the optimization variables in the direction of the negative gradient of the Lagrangian while enforcing feasibility for $g_i(x)$ and $h_j(x)$:
\begin{align}
\label{eq:circuit_dynamics}
\diff{x}{t}= - \gamma \nabla_x \mathcal{L}(x,\lambda,\mu),
\end{align}
where $\gamma > 0$ is a scalar parameter. Each optimization variable $x_k$ is represented by a node voltage $v_k$, and its value at any time is given by the voltage at that node. We illustrate this process with a circuit example in Section~\ref{sec:Circuit Example}. For more details on the synthesis of these circuits for optimization problems, we refer readers to prior work~\cite{Kennedy_1988, Costantini_2008, Poon_2022, Sawant_2024,Sawant_2025}.

At steady state, pertinent circuit voltages $v$ correspond to the optimization variables $x$, and together with the dual variables $\lambda$, $\mu$  satisfy the Karush–Kuhn–Tucker (KKT) conditions, which are first-order necessary conditions for optimality when constraint qualification holds~\cite[Chapter~12]{Nocedal_2006}: 
\begin{equation}
\label{eq:KKT_conditions}
\begin{aligned}
\nabla_x \mathcal{L}(x^\star,\lambda^\star,\mu^\star) &= 0,  \quad \mathrm{(Stationarity)} \\
g(x^\star)  \geq 0, \;
h(x^\star)&= 0, \quad \mathrm{(Primal\,feasibility)}\\
\lambda^\star & \leq 0, \quad \mathrm{(Dual\,feasibility)}\\
\lambda^\star \circ g(x^\star) &= 0, \quad \mathrm{(Complementary\,slackness)}
\end{aligned}
\end{equation} where $\circ$ denotes the element-wise product. The solution $x^\star \in \mathbb{R}^N$is globally optimal when the optimization problem~\eqref{eq:OP} is convex, while in the nonconvex case it represents a KKT stationary point.

With any network involving dynamics of the form we consider, stability is an important consideration. These dynamics can be analyzed using control-theoretic tools such as Lyapunov stability theory and the Kalman-Yakubovich-Popov~(KYP) lemma, to establish conditions for global asymptotic and exponential stability~\cite{Kennedy_1988,Forti_1995,Costantini_2008,Adegbege_2021,Sawant_2025}.

\subsection{Analog Circuit Example}
\label{sec:Circuit Example}
\begin{figure}[t!]
\centering
\includegraphics[width=\linewidth]{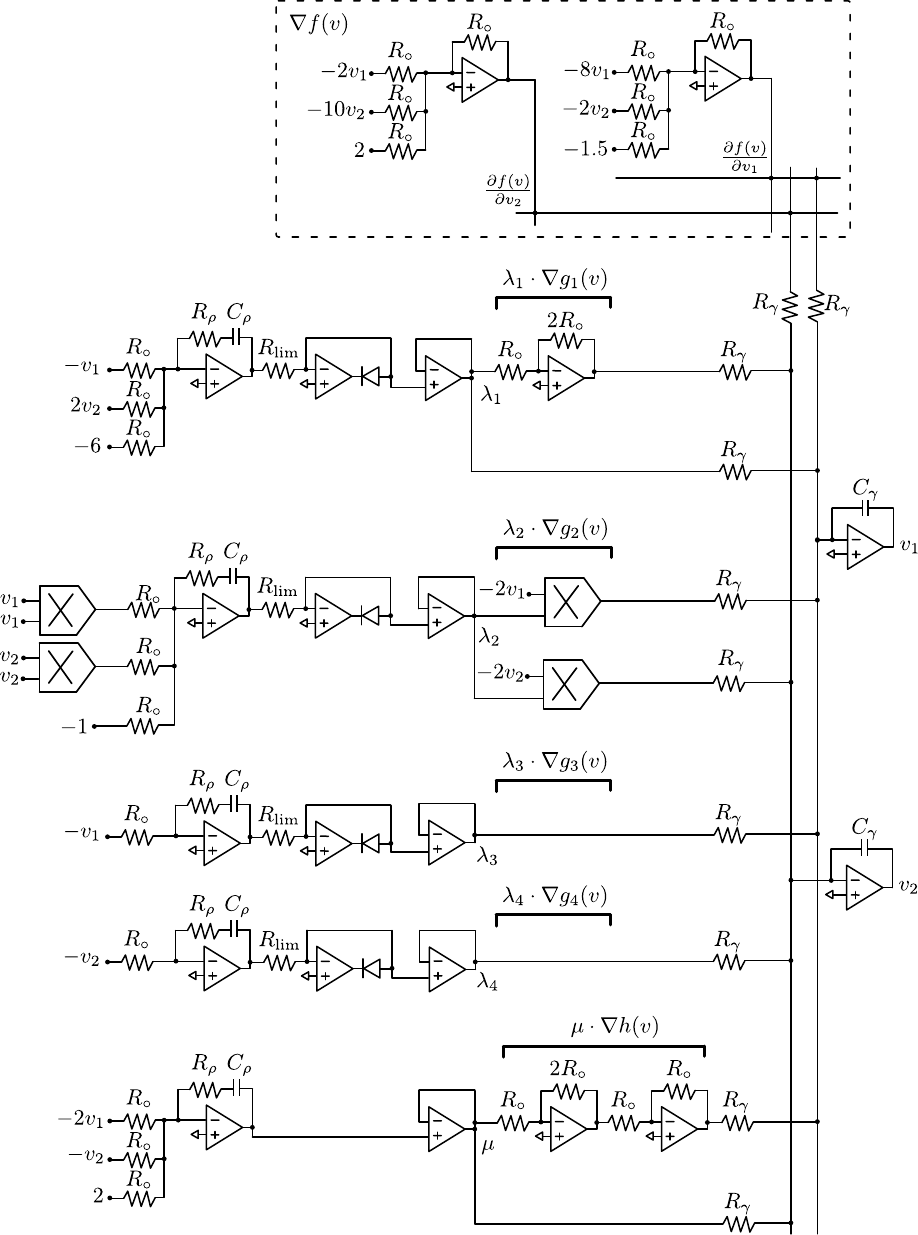}   
\caption{Circuit schematic for the augmented Lagrangian method to solve the optimization problem in~\eqref{eq:circuit_example}. Similarly, the primal-dual and penalty methods can be implemented by short-circuiting $R_\rho$ and $C_\rho$, respectively. Component values are $R_\circ = R_{\mathrm{lim}} = R_\gamma = 10\ \mathrm{k}\Omega$, $R_\rho = 100\ \mathrm{k}\Omega$, $C_\rho = 10\ \mathrm{nF}$, and $C_\gamma = 100\ \mathrm{nF}$.}
\label{fig:example_circuit}
\end{figure}

To illustrate mapping optimization problems to analog circuits, we consider the following optimization problem based on~\cite{Maros_1999}:
\begin{subequations}
\label{eq:circuit_example}
\begin{align}
\underset{x \in \mathbb{R}^2}{\mathrm{minimize}} \quad f(x)= \, &\frac{1}{2}(8x_1^2 + 4x_1x_2 + 10x_2^2) \nonumber\\
& \quad + 1.5x_1 - 2x_2 + 4, \\
\mathrm{subject\,\,to} \quad g(x): \quad  & x_1 - 2x_2+6 \geq 0, \\
 &\hspace{2mm}1-x_1^2-x_2^2 \geq 0, \nonumber \\
 &\hspace{16mm} x_1 \geq 0, \nonumber \\
 &\hspace{16mm}x_2 \geq 0, \nonumber \\
h(x): \quad &\hspace{0.5mm}2x_1 + x_2 -2= 0.
\end{align}
\end{subequations}

Figure~\ref{fig:example_circuit} shows the analog realization of~\eqref{eq:circuit_example}, adopted from the analog circuit in~\cite{Kennedy_1988,Sawant_2025}. The node voltages $v_1$ and $v_2$ correspond to the optimization variables $x_1$ and $x_2$. Operational amplifier (op-amp) stages labeled $\nabla f(v)$, $\nabla g_i(v)$, and $\nabla h_j(v)$ generate the partial derivatives of the objective function, the $i$-th inequality constraints, and $j$-th equality constraints, respectively. For the problem in \eqref{eq:circuit_example}, the gradients are: $\nabla f(v)=\begin{bmatrix} 8v_1 + 2v_2 + 1.5 & 2v_1 + 10v_2 - 2 \end{bmatrix}^\top, \nabla g_1(v) = \begin{bmatrix} 1 & -2 \end{bmatrix}^\top, \nabla g_2(v) = \begin{bmatrix}-2v_1 & -2v_2 \end{bmatrix}^\top, \nabla g_3(v) = \begin{bmatrix} 1 & 0\end{bmatrix}^\top, \nabla g_4(v) = \begin{bmatrix} 0 & 1 \end{bmatrix}^\top$, and $\nabla h(v) = \begin{bmatrix} 2 & 1 \end{bmatrix}^\top$.

The dual variables $\lambda_i$ for the inequality constraints are implemented as auxiliary nodes. An op-amp stage with diode feedback effectively applies the element-wise $\min(0, z)$ operation, yielding zero when the inequality constraint is satisfied ($g_i(v) \geq 0$) and otherwise passing its input $z$. The dual variables $\mu_j$ for the equality constraints are realized in the same way, but without the op-amp stage having diode feedback, since the optimal solution must satisfy the equality constraint exactly. The dynamics of these dual variables are governed by the op-amp stage containing $R_\rho$ and $C_\rho$: with both $R_\rho$ and $C_\rho$ present, the circuit implements an Augmented Lagrangian method, short-circuiting $R_\rho$ yields a Primal-dual method, while short-circuiting $C_\rho$ recovers a Penalty method~\cite{Sawant_2025}. 

The op-amp stages configured as integrators, with capacitors $C_\gamma$ connected to $v_1$ and $v_2$, integrate the net currents generated by the circuit blocks implementing $\nabla f(v)$, $\nabla g_i(v)$, and $\nabla h_j(v)$, realizing the dynamics in~\eqref{eq:circuit_dynamics}:
\begin{align*}
\diff{v}{t}= - \frac{1}{R_\gamma C_\gamma} \nabla_x \mathcal{L}(v,\lambda,\mu),
\end{align*}
where $v=\begin{bmatrix} v_1 & v_2 \end{bmatrix}^\top$.

At equilibrium, the input currents to the op-amp integrators vanish, causing the capacitor $C_\gamma$ voltages to remain constant. In steady-state, the circuit variables, $v$, together with the dual variables, $\lambda$ and $\mu$, satisfy the KKT conditions in~\eqref{eq:KKT_conditions}. Thus, the steady-state voltages $v_1$ and $v_2$ correspond to the globally optimal solution of the optimization problem~\eqref{eq:circuit_example}.

\section{Toolchain Overview}
\label{sec:Toolchain Overview}
\begin{figure}[t!]
\centering
\includegraphics[width=\linewidth]{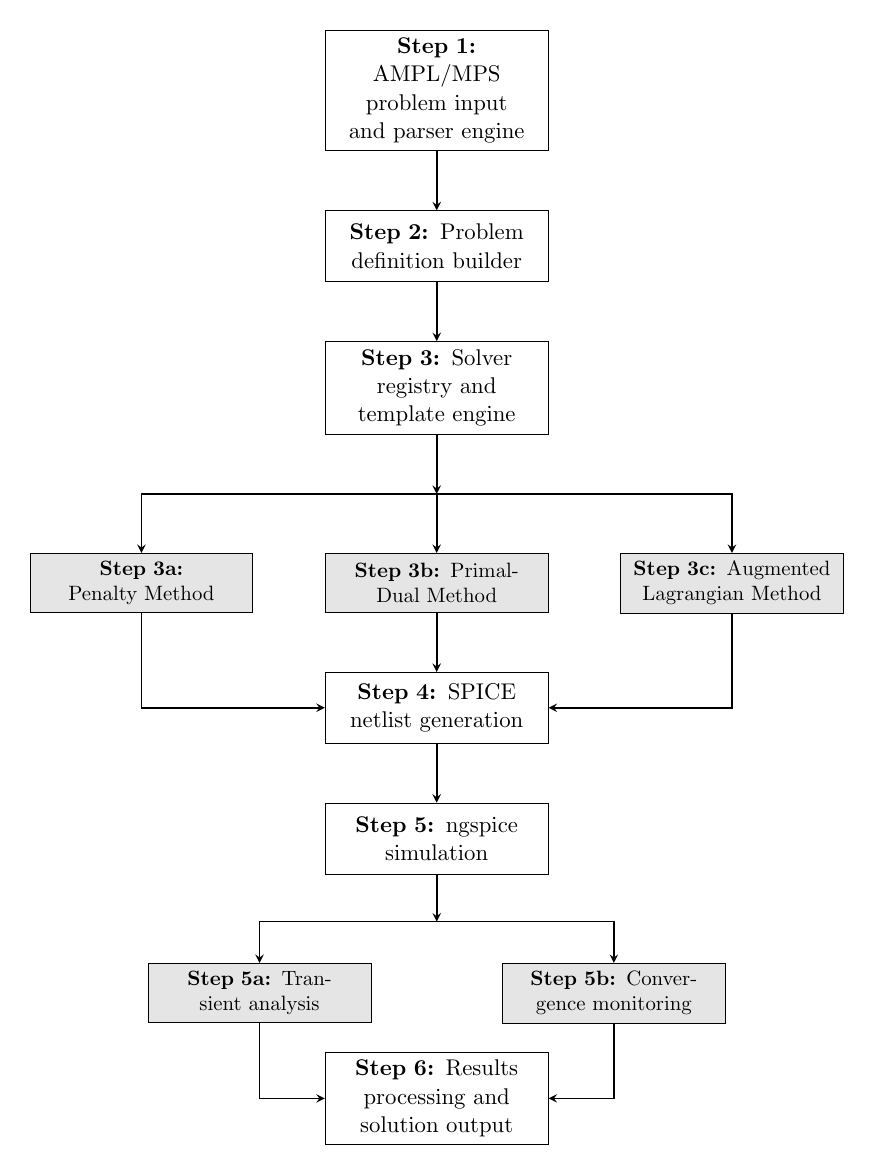}
\caption{Overview of the automated software toolchain flow.}
\label{fig:flow}
\end{figure}

This section presents the overall flow of the proposed toolchain. Figure~\ref{fig:flow} illustrates the general flow: the user provides a high-level problem description in a standard optimization format (AMPL/MPS). The toolchain then automatically synthesizes an analog circuit netlist. The process is implemented in two steps as summarized below.

\subsection{Problem Translation and Circuit Synthesis}
The toolchain first parses the input file and converts the mathematical expressions into an internal symbolic representation. It then automatically synthesizes a SPICE netlist by mapping these symbolic elements to circuit templates from a predefined library. This stage performs the translation from an abstract optimization problem to a circuit description. The output is a text-based netlist that specifies components (e.g., op-amps, resistors, and multipliers) and their interconnections.

While the \texttt{ngspice} circuit simulator is used for verification, the synthesized circuits are intended for physical realization rather than simulation alone. The SPICE netlist serves as an implementation-ready specification that can be mapped to a schematic and layout using standard analog building blocks (e.g., op-amps, resistors, and capacitors) and subsequently prototyped on a printed circuit board (PCB) or as an application-specific integrated circuit (ASIC).

Standard electronic components such as op-amps, diodes, multipliers, resistors, and capacitors suffice to realize quadratic objective functions with linear and quadratic constraints. To ensure scalability to general nonlinear optimization problems, complex functions are abstracted in simulation via behavioral voltage sources that support arbitrary mathematical expressions (e.g., \texttt{sin}). These sources can be replaced by dedicated analog macro-circuits (e.g., translinear multipliers or switched-capacitor circuits) if the synthesized circuit is intended to be fully analog, or by hybrid analog-digital circuit elements (e.g., digital-to-analog converter (DAC)/analog-to-digital converter (ADC) interfaces or microcontrollers)~when a mixed-signal implementation is desired~\cite{Poon_2022,Sawant_2024}.

\subsection{Simulation and Verification}
The generated netlist is passed to an industry-standard SPICE simulator~(\texttt{ngspice}). A transient analysis is performed, allowing the circuit's voltages, which represent the optimization variables, to evolve and settle at an equilibrium point that satisfies the KKT conditions~\eqref{eq:KKT_conditions}. The final steady-state voltages, which represent the optimal values of the optimization variables, are then extracted and reported as the solution. The convergence time obtained from the simulation also provides an estimate of the solution time for the corresponding physical analog or hybrid (analog/digital) circuits.

The analog optimization toolchain consists of the following software modules, shown in Fig.~\ref{fig:flow}, which illustrate the steps for analog circuit synthesis:   
\begin{enumerate}
\item \textbf{Parser Engine (Step 1).}
Extracts the optimization problem from the \textit{AMPL/MPS file format} and processes it to obtain variables, constraints, and objective functions. Gradient information is generated via automatic differentiation, with hierarchical caching and multithreaded pipelines.

\item \textbf{Problem Definition Builder (Step 2).}
Transforms the parsed problem into a structured intermediate representation for the subsequent template engine. Symbolic preprocessing, powered by a customized \texttt{SymPy}-based backend, computes exact constraint Jacobians and objective gradients, ensuring solver-ready formulation.

\item \textbf{Solver Registry and Methods (Step 3, 3a--3c).}  
Includes several optimization methods that can be selected:~\cite{Sawant_2025} 
\begin{itemize}
    \item \textit{Penalty Method (Step 3a)}: Uses large gains as penalties for constraint enforcement; a steady-state error occurs when the optimal solution lies on an active constraint (fewest analog circuit components, steady-state error).
    \item \textit{Primal-Dual Method (Step 3b)}: Incorporates dual-variable dynamics to ensure convergence to the optimal solution without steady-state error (no steady-state error, slower convergence).
    \item \textit{Augmented Lagrangian (Step 3c)}: Combines penalty terms and dual-variable dynamics to achieve faster convergence to the optimal solution without steady-state error (faster convergence without steady-state error, more analog circuit components).
\end{itemize}

\item \textbf{Netlist Generation (Step 4).}  
Selected solver formulations are translated into SPICE-compatible analog netlists through a \texttt{Jinja2}-driven template synthesis framework, as outlined in Algorithm~\ref{alg:analog_synthesis}. This process provides a link between symbolic optimization models and circuit-level simulation, ensuring that mathematical abstractions are rendered into physically implementable analog circuits.

\item \textbf{Simulation and Analysis (Steps 5--5b).}  
Netlists are simulated using \texttt{ngspice}, including transient analysis (Step 5a) and convergence monitoring (Step 5b).

\item \textbf{Results Processing and Solution Output (Step 6).}  
Simulation outputs are processed to extract variable values, compute convergence metrics, and produce validated solutions for the user.
\end{enumerate}

\subsection{Automated Synthesis Algorithm}
\begin{algorithm}[t!]
\caption{Analog Netlist Synthesis}
\label{alg:analog_synthesis}
\begin{algorithmic}[1]
\Require Objective function $f(x)$, optimization variables $\{x_j\}$, constraints $\{g_i(x)\}$, $\{h_j(x)\}$, method selection $\mathcal{M} \in \{\text{Penalty, Primal-Dual, Augmented Lagrangian}\}$
\Ensure SPICE Netlist \(N_{\text{spice}}\)

\State Initialize template $T$, component list $C \gets \emptyset$

\For{each variable \(x_j\)}
    \State Instantiate an inverting op-amp integrator with $(R_\gamma, C_\gamma)$
    \State Assign node voltage $v_j \leftrightarrow x_j$
\EndFor

\For{each gradient term \(\partial f/\partial x_j\)}
    \State Realize as a behavioral voltage source feeding the input summing node of the corresponding integrator
\EndFor

\For{each constraint expression}
    \State Accumulate the weighted terms of each constraint function $g_i(x)$ (or equality $h_j(x)$) at the input of an op-amp stage
\EndFor

\For{each inequality constraint $g_i(x) \geq 0$}
    \If{$\mathcal{M} = \text{Penalty}$}
        \State Configure the op-amp stage as an inverting amplifier with $R_\rho$ as the feedback resistor to set the penalty
    \ElsIf{$\mathcal{M} = \text{Primal-Dual}$}
        \State Configure the op-amp stage as an inverting integrator with capacitor $C_\rho$
        
    \ElsIf{$\mathcal{M} = \text{Augmented Lagrangian}$}
        \State Configure the op-amp stage as an inverting stage, using a series combination of $R_\rho$ and $C_\rho$ in the feedback
    \EndIf
\EndFor

\For{each inequality constraint $g_i(x)\geq 0$}
    \State Connect the output of the method-specific op-amp to an inverting op-amp with diode feedback
    \State Feedback activated only when $g_i(x) < 0$
\EndFor

\For{each equality constraint $h_j(x)=0$}
    \State Omit the inverting op-amp stage with diode feedback
\EndFor

\For{each constraint term in $g_i(x)$ or $h_j(x)$}
    \If{term is linear $c_{ij}x_j$}
        \State Instantiate inverting op-amp amplifier with gain $c_{ij}$
    \Else
        \State Instantiate a multiplier or other appropriate nonlinear circuit
    \EndIf
\EndFor

\State Assemble all components $C$ into $T$, generate $N_{\text{spice}}$
\State \Return $N_{\text{spice}}$
\end{algorithmic}
\end{algorithm}

We now present Algorithm~\ref{alg:analog_synthesis}, which synthesizes analog circuits for solving constrained optimization problems. The synthesis of analog optimization circuits maps the optimization problem into a SPICE-compatible netlist. The optimization variables \(x_j\) are realized as node voltages \(v_j\), the objective function gradient \(\nabla f(x)\) is implemented using behavioral voltage sources, and the constraints \(\{g_i(x), h_j(x)\}\) are enforced through analog circuits, as illustrated in Fig.~\ref{fig:example_circuit}.

\section{Numerical Results} \label{sec:numerical_results}
We evaluated the proposed toolchain on a set of benchmark problems to assess its performance. The simulations were performed on an Apple M3 Pro system with 16 GB of RAM, using \texttt{ngspice} v41 as the simulation backend. This study uses a benchmark set available at~\cite{benchmarking_files}, based on standard AMPL model files~\cite{plato_ampl_nlp}.

We focused on convex problems to ensure that a global optimum exists, providing a fair basis for comparing the solution and computational speed of different solvers. The performance of the synthesized analog circuits was evaluated against IPOPT~\cite{IPOPT}, a state-of-the-art interior-point solver, to provide a baseline for comparison with digital solvers.

The primary performance metric for the analog solver is the convergence time from the SPICE transient analysis simulation (`.TRAN' analysis). This simulation time is used as a proxy for the physical settling time of a corresponding hardware implementation. For each problem, the total solution time for the analog circuit was recorded from the start of the transient analysis until the node voltages representing the optimization variables reached a steady state (settling to within $0.01\%$ of their final value). Digital solver times were measured using the CPU time reported by IPOPT within the AMPL environment~\cite{AMPL}. 

\begin{figure}[t!]
\centering
\includegraphics[width=\linewidth]{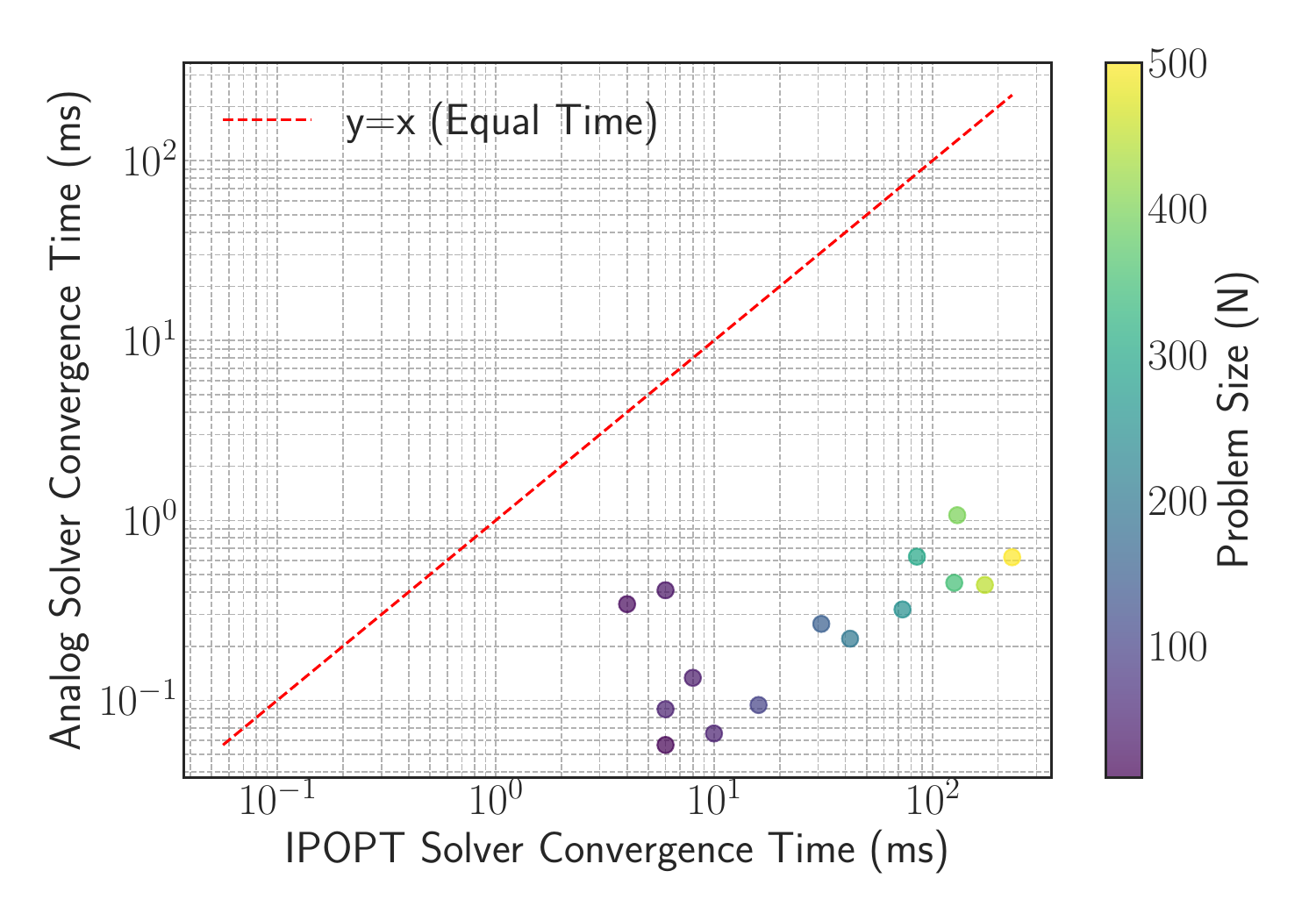}
\caption{Analog solver speed comparison with IPOPT solver.}
\label{fig:solver_comparison}
\end{figure}

Figure~\ref{fig:solver_comparison} compares the computation time in milliseconds (ms) of the analog solver and IPOPT across varying problem sizes~(number of optimization variables, $N$). The logarithmic scale shows that, as problem complexity increases, the performance gap widens in favor of the analog solver. While IPOPT's solution times exhibit the characteristic polynomial growth expected from interior-point methods, the analog solver maintains relatively consistent convergence times due to its inherent parallel operation. For problems with fewer than $100$ variables, both methods perform comparably; however, beyond this threshold, the analog solver's advantages become evident. For example, on a problem with $500$ variables, $50$ linear constraints, and $50$ quadratic constraints, the synthesized analog circuit converged to a solution in $0.62$~ms, whereas the IPOPT solver required $232$~ms for the same problem---a speedup of approximately $370\times$.

\begin{figure}[t!]
\centering
\includegraphics[width=\linewidth]{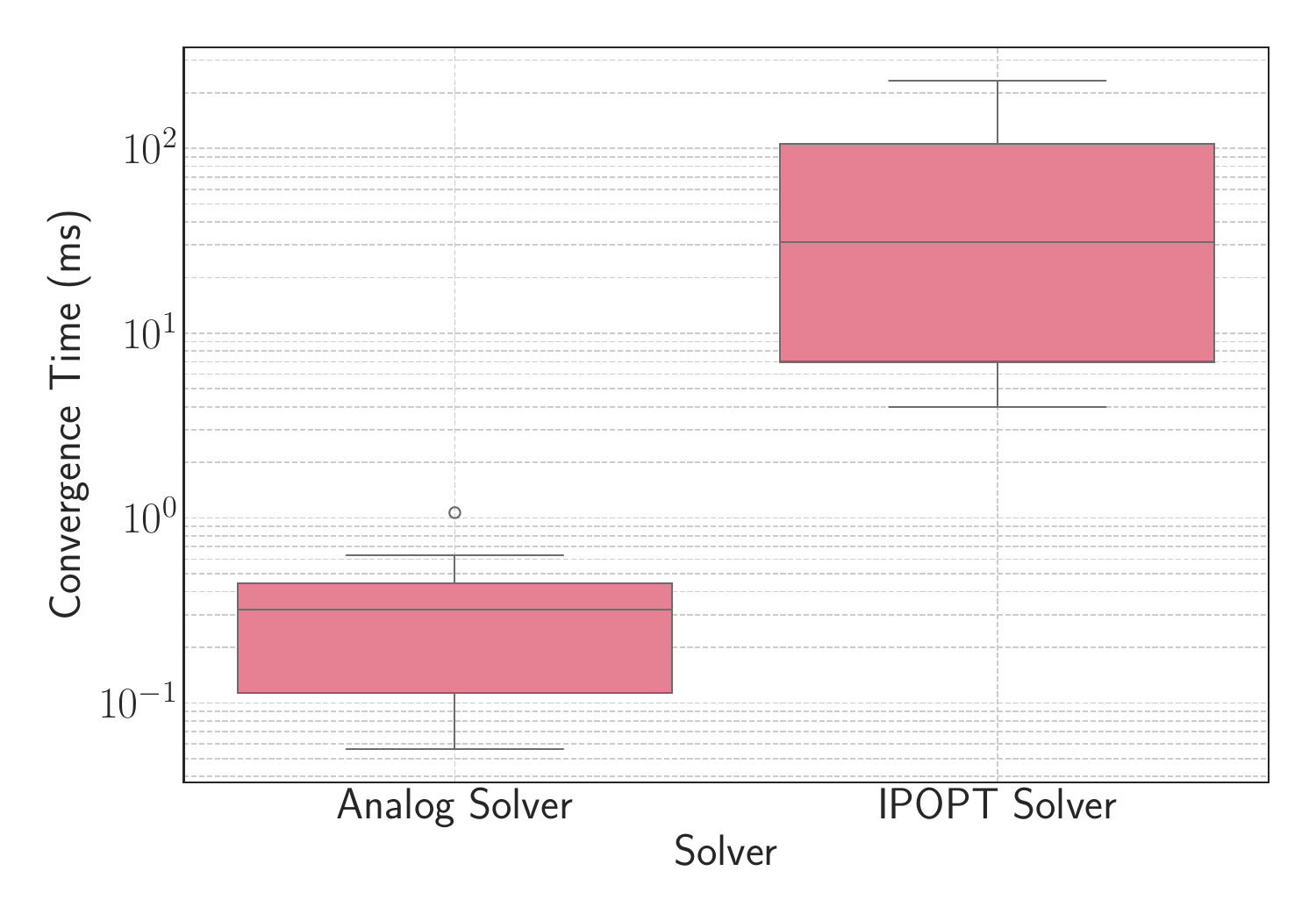}
\caption{Distribution of convergence time of analog solver and IPOPT solver.}
\label{fig:time_distribution}
\end{figure}

Figure~\ref{fig:time_distribution} shows the distribution of convergence times, using box plots to illustrate the consistency of both approaches. The analog solver exhibits a tight distribution with minimal variance, indicating consistent performance across different problem instances. The median convergence time of $0.32$~ms for the analog solver contrasts with IPOPT's median of $31$~ms, representing a $97\times$ improvement. The narrow interquartile range of the analog solver indicates timing predictability, which is important for real-time applications.

This speedup was achieved without loss of solution accuracy. For the $500$-variable problem, the objective value obtained by the analog circuit ($-1489.9999$) deviated from the IPOPT solution ($-1489.998$) by less than $0.0001\%$, confirming that the circuit's steady-state equilibrium point corresponds to the problem's mathematical optimum. This performance advantage stems from the inherent parallelism of the analog solver. The circuit's continuous-time dynamics, which map directly to the KKT optimality conditions, allow all state variables to evolve simultaneously towards the solution. 

\begin{figure}[t]
\centering
\includegraphics[width=\linewidth]{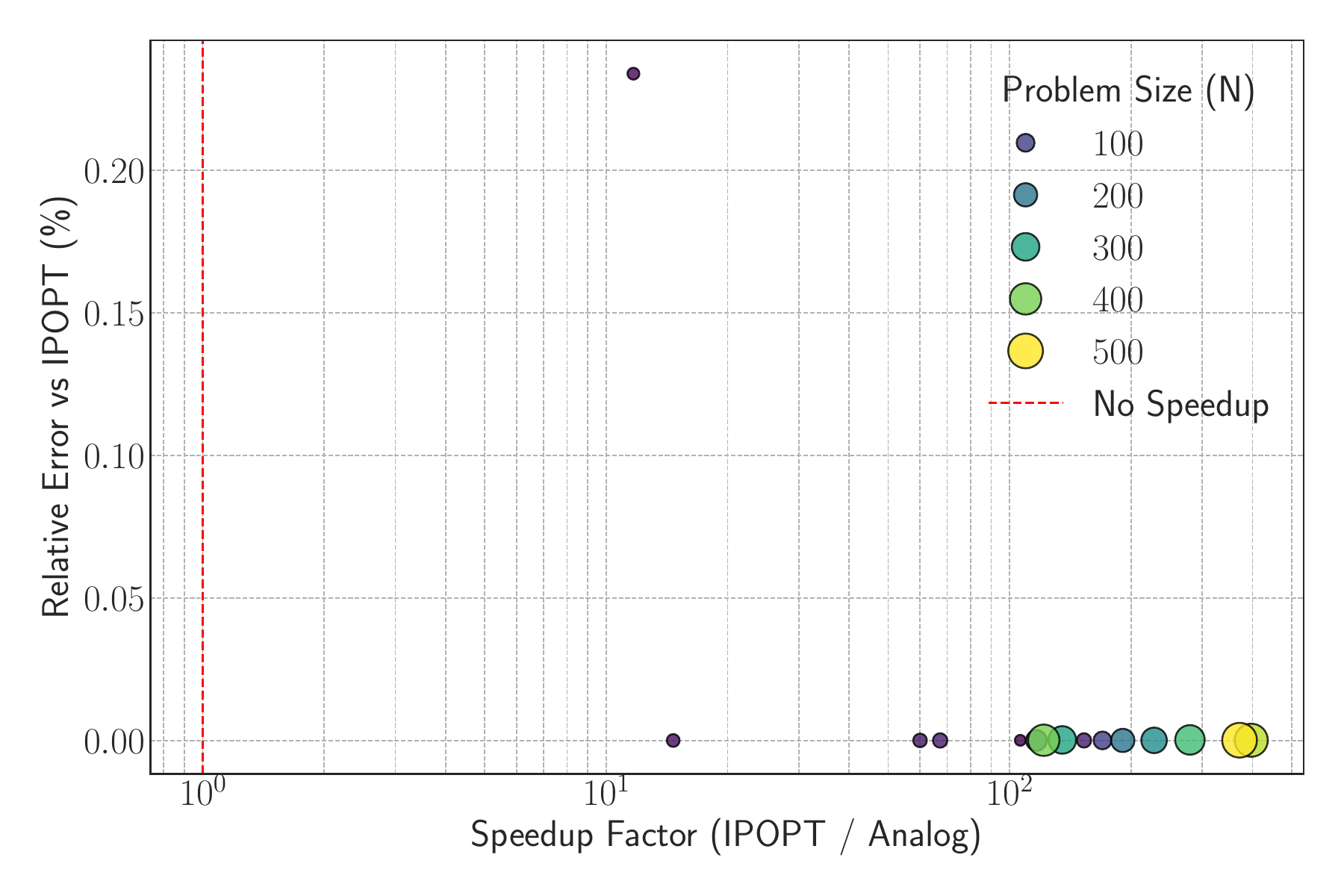}
\caption{Analog solver speed-accuracy trade-off with IPOPT solver.}
\label{fig:speed_accuracy}
\end{figure}

Figure~\ref{fig:speed_accuracy} examines the trade-off between solution speed and accuracy. The analog solver achieves sub-millisecond convergence times while maintaining solution accuracy within $0.02\%$ of the optimal values found by IPOPT (the accuracy of the analog solution is subject to component tolerances). The clustering of analog solver results in the lower-right region of the plot (fast convergence, high accuracy) underscores its suitability for applications requiring both speed and precision. This contrasts with the iterative nature of interior-point methods, which at each iteration require solving a large system of linear equations and performing matrix factorizations, a computationally expensive process that scales poorly with problem size. Table~\ref{tab:solver_comparison} summarizes the comparison: the analog solver achieved a mean speedup of $161.56\times$ and a maximum of $397.2\times$, while maintaining a mean relative error of $0.02\%$ and a median of $0.00\%$, indicating that, for most problems, the solution obtained from the analog solver is indistinguishable from the IPOPT solution.

\begin{table}[b!]
\centering
\caption{Comparison of Analog Solver and IPOPT Solver Performance.}
\label{tab:solver_comparison}
\renewcommand{\arraystretch}{1.3}
\setlength{\tabcolsep}{12pt}
\begin{tabular}{l c c}
\toprule
\midrule
\textbf{Metric} & \textbf{Analog Solver} & \textbf{IPOPT Solver} \\
\midrule
Mean Time (ms)            & $0.35$   & $63.27$ \\
Median Time (ms)          & $0.32$   & $31.00$ \\
Maximum Speedup           & $397.2\times$ & - \\
Mean Speedup              & $161.56\times$ & - \\
Median Speedup            & $135.09\times$ & - \\
Mean Relative Error (\%)   & $0.02$   & - \\
Median Relative Error (\%) & $0.00$   & - \\
\bottomrule
\end{tabular}
\end{table}

During the scalability analysis, we found that the main limitation was not in the synthesis flow, but in the computational limits of the \texttt{ngspice} simulator when handling the matrix factorization of very large netlists. Simulating a parallel analog circuit on a sequential digital computer requires solving a large matrix of circuit equations at each time step. The computational cost of this step grows nonlinearly with the number of circuit nodes and the density of their interconnections. In practice, the point at which the simulation becomes prohibitive also depends on the hardware used to run the simulator. For example, on an Apple MacBook M3 Pro with $16$\, GB of memory, simulation typically halts once the circuit reaches about $10^6$ nodes. At this scale, representative problem instances include:
\begin{itemize}
\item \textbf{Sparse case:} Roughly $ >10{,}000$ optimization variables with $>10{,}000$ linear constraints, each involving about $1{,}000$ variables. 
\item \textbf{Dense case:} Approximately $>500$ optimization variables with $>300$ quadratic constraints and $>200$ linear constraints, each coupling nearly all variables.
\end{itemize}
Thus, both problem density and the total number of variables and constraints determine scalability limits. These observations suggest that the synthesis methodology scales with problem size; however, the performance ceiling is imposed by the sequential nature of the verification tool. A physical analog solver implementation would not encounter this bottleneck and could handle large-scale optimization problems in real-time.

\section{Conclusions \& Future Work}
\label{sec:conclusions}
This paper presents an automated software toolchain for synthesizing and verifying analog circuits that solve constrained optimization problems. The toolchain demonstrates scalability by automatically generating SPICE netlists that can handle problems with thousands of variables, thereby overcoming the limitations of manual, hardware-based implementations, which are limited to a few optimization variables. This software toolchain provides a foundation for the systematic design and prototyping of analog circuits to solve optimization problems of varying complexity for real-time optimization applications. Future work will focus on extending the toolchain to distributed simulation environments such as \texttt{Xyce}, and on advancing automated mapping strategies for deployment to physical hardware platforms, including FPAAs.
\bibliographystyle{IEEEtran}
\bibliography{references}

\end{document}